\documentclass[runningheads]{svmult}

\usepackage{graphicx}  
\usepackage{subeqnar}  
\usepackage{multicol}  
\usepackage{physprbb}  
\makeindex             

\begin{document}
\title*{The Virtual Observatory as a Tool to Study Star Cluster
Populations in Starburst Galaxies}
\toctitle{The Virtual Observatory as a Tool to Study Star Cluster Populations
\protect\newline in Starburst Galaxies}
%
%
\titlerunning{Use of the Virtual Observatory to Study Star Cluster Systems}
%
\author{Richard de Grijs\inst{1}
\and Uta Fritze--v. Alvensleben\inst{2}
}
\authorrunning{R. de Grijs \& U. Fritze--v. Alvensleben}
%
%
\institute{Institute of Astronomy, University of Cambridge, Madingley
Road, Cambridge CB3 0HA, U.K.
\and University of G\"ottingen, Universit\"atssternwarte,
Geismarlandstra{\ss}e 11, 37083 G\"ottingen, Germany}

\maketitle              


\section{The Cluster Luminosity Function}

The cluster luminosity function (CLF) is one of the most important
diagnostics in the study of old globular and young compact star cluster
populations.  While the CLFs of young star cluster systems are closely
approximated by power laws, as predicted by standard cluster formation
models (e.g., \cite{degrijs.ref2} \cite{degrijs.ref4}), for old globular
cluster systems the shape of the CLF, and also of the {\it initial} mass
function, is well-established to be roughly Gaussian, with an apparently
universal turnover luminosity (mass). 

Processes responsible for the transformation of the CLF of young star
cluster systems into the lognormal distributions characteristic of older
CLFs include differential fading due to aging of the cluster population,
the preferential depletion of low-mass clusters by evaporation and tidal
interaction with the gravitational field of the parent galaxy, and the
preferential destruction of high-masss clusters by dynamical friction. 
The relative importance of these destruction processes is still
controversial (see, e.g., \cite{degrijs.ref4} \cite{degrijs.ref5}
\cite{degrijs.ref6}). 

One should be cautious, however, to generalize any of the ``standard''
models, since they only apply to quiescent, non-interacting galaxies. 
In the high-pressure environments and the time-varying potentials of
interacting and merging galaxies we expect significant differences for
the cluster formation and destruction processes, mass spectra, and
time-scales for the dynamical evolution of star clusters due to, e.g.,
external pressure and shocks. 

\section{Scientific Questions and the Virtual Observatory}

Using the wealth of archival data from the {\sl Hubble Space Telescope}
and large ground-based facilities, we are currently using the ST-ECF
ASTROVIRTEL tools\footnote{E.g., {\it querator}
(http://archive.eso.org/querator) for imaging observations and {\it
listator} (http://archive.eso.org/listator) for spectroscopic
observations.} to obtain CLFs and colour distributions in several
optical and/or near-infrared passbands, in order to address these issues
statistically. 

\begin{figure}[h]
\begin{center}
\includegraphics[width=8cm]{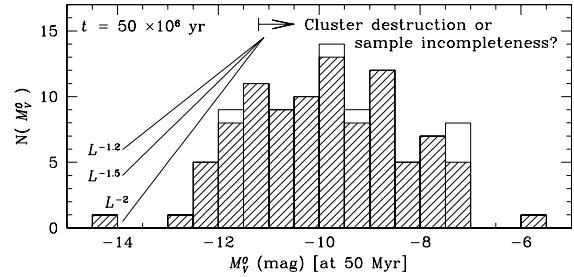}
\end{center}
\vspace{-4cm}
\caption[]{CLF for the cluster sample in M82's fossil starburst region
``B'', corrected individually to a fiducial age of 50 Myr
\cite{degrijs.ref1}.  The CLF may be broadened due to dynamical
destruction of lower mass clusters, although the degree to which our
cluster sample is incomplete for these brightnesses is uncertain. 
Questions like these can now be addressed with the large amount of data
available in the archives.}
\label{degrijs1.fig}
\end{figure}

We caution strongly that it is obviously very important to obtain
accurate ages for the individual clusters and to correct the
observational CLF to a common age before interpreting the ``intrinsic''
CLF, see Fig.  \ref{degrijs1.fig} \cite{degrijs.ref1}
\cite{degrijs.ref3}. 

The ASTROVIRTEL initiative allows us, for the first time, to address
questions such as:

\begin{enumerate}

\item Is the {\it intrinsic} CLF (or, alternatively, the cluster initial
mass function) a power law in all cases?

\item What are the time-scales for the evolution from the power-law
young compact CLFs into old globular CLFs?

\item Is there a clear difference between OB associations/open clusters
and compact globular clusters at formation or do they all form a
continuum mass spectrum?

\item Can we ``observe'' dynamical cluster destruction by comparison of
young star cluster systems in ongoing mergers and merger remnants of
various ages?

\item Is there an environmental dependence on the efficiency and
therefore on the time-scale of low-mass star cluster depletion?

\item Is the (old) globular CLF {\it universal}? To what extent and how
reliably can it be used as a (secondary) distance indicator and to
estimate $H_0$?

\item Do stronger starbursts favour the formation of more massive
clusters?

\item Does the ratio between field star formation and star formation in
clusters depend on the galaxy or interaction properties?

\item Can we determine the violent star and cluster formation histories
of galaxies based on star cluster colour distributions and deep
luminosity functions?
  
\end{enumerate}

\end{document}